\documentclass[showpacs,prl,twocolumn,floatfix]{revtex4}
\usepackage{graphicx, epsfig,amsmath, amssymb, subfigure}

\begin{document}

\title{Self-organization versus top-down planning in the evolution of a city}

\author{Marc Barthelemy$^{1,2}$ and Patricia Bordin$^{3,4}$ and Henri
  Berestycki$^{2}$ and Maurizio Gribaudi$^{5}$}

\affiliation{$(1)$ Institut de Physique Th\'eorique, CEA, CNRS-URA 2306, F-91191,
  Gif-sur-Yvette, France.}
\affiliation{$(2)$ Centre d'Analyse et de Math\'ematiques Sociales, EHESS, 190-198 avenue
  de France, 75244 Paris, France.}
\affiliation{$(3)$ Universit\'e Paris-Est, Institut de Recherche en
  Constructibilit\'e, ESTP, F-94230, Cachan, France.}
\affiliation{$(4)$ Universit\'e Paris Diderot, Sorbonne Paris Cit\'e, Institut des Energies
de Demain (IED), 75205 Paris, France.}
\affiliation{$(5)$ Laboratoire de D\'emographie et Histoire Sociale, EHESS, 190-198 avenue de
France, 75244 Paris, France.}

\begin{abstract}

  Interventions of central, top-down planning are serious limitations
  to the possibility of modelling the dynamics of cities. An example
  is the city of Paris (France), which during the 19th century
  experienced large modifications supervised by a central authority,
  the `Haussmann period'. In this article, we report an empirical
  analysis of more than 200 years (1789-2010) of the evolution of the
  street network of Paris. We show that the usual network measures
  display a smooth behavior and that the most important quantitative
  signatures of central planning is the spatial reorganization of
  centrality and the modification of the block shape
  distribution. Such effects can only be obtained by structural
  modifications at a large-scale level, with the creation of new roads
  not constrained by the existing geometry. The evolution of a city
  thus seems to result from the superimposition of continuous, local
  growth processes and punctual changes operating at large spatial
  scales.

\end{abstract}

\maketitle



\section*{Introduction}

A city is a highly complex system where a large number of agents
interact, leading to a dynamics seemingly difficult to understand.
Many studies in history, geography, spatial economics, sociology, or
physics discuss various facets of the evolution of the city
\cite{Mumford,Batty,Angel:book,Fujita:2001,Glaeser:2011,Makse:1995,Bettencourt:2007,Marshall:2009,Batty:2009,Geddes:1949}. From
a very general perspective, the large number and the diversity of
agents operating simultaneously in a city suggest the intriguing
possibility that cities are an emergent phenomenon ruled by
self-organization \cite{Batty}. On the other hand, the existence of
central planning interventions might minimize the importance of
self-organization in the course of evolution of cities. Central
planning --here understood as a top-down process controlled by a
central authority -- plays an important role in the city, leaving long
standing traces, even if the time horizon of planners is limited and
much smaller than the age of the city. One is thus confronted with the
question of the possiblity of modelling a city and its expansion as a
self-organized phenomenon. Indeed central planning could be thought of
as an external perturbation, as if it were foreign to the
self-organized development of a city. The recent digitization and
georeferentiation of old maps will enable us to test quantitatively
this effect, at least at the level of the structure of the road
network. Such a transportation network is a crucial ingredient in
cities as it allows individuals to work, transport and exchange goods,
etc., and the evolution of this network reflects the evolution of the
population and activity densities \cite{Southworth,Xie:2009}.  These
network aspects were first studied in the 1960s in quantitative
geography \cite{Haggett:1969}, and in the last decade, complex
networks theory has provided significant contributions to the
quantitative characterization of urban street
patterns~\cite{Jiang:2004,Porta:2006,Lammer:2006,Crucitti:2006,Marshall:2006,Cardillo:2006,Xie:2007,Barthelemy:2008,Courtat:2011,Barthelemy:2011,Strano:2012}.

\begin{figure*}
\centering
	\includegraphics[scale = 0.50]{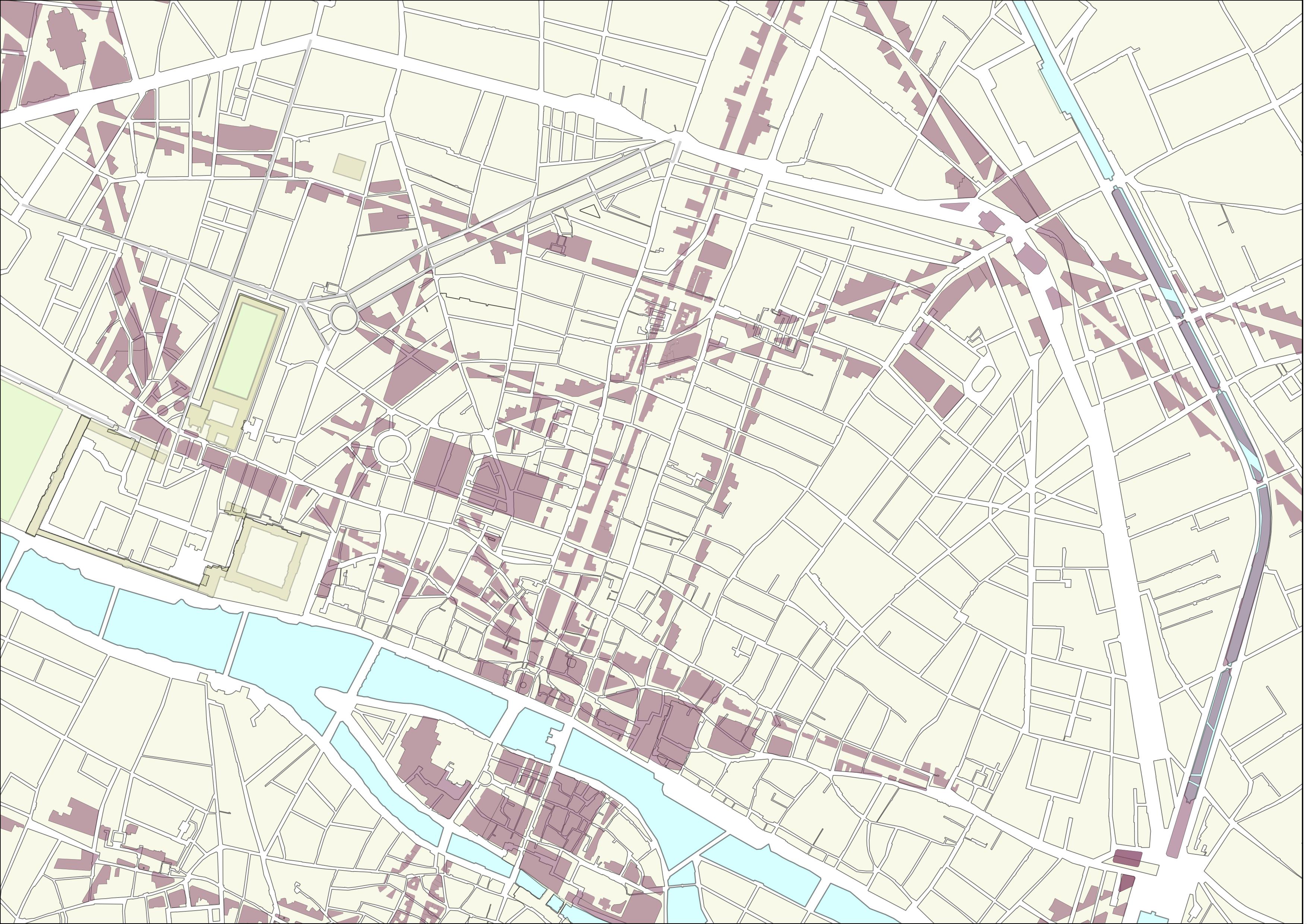}
	\caption{Illustration on a small area of the impact of
          Haussmann's transformations. On the yellow background, we
          show the parcel distribution before Haussmann (extracted
          from the Vasserot cadastre, 1808-1836), and in brown we show
          the new buildings delineating the new streets as designed by
          Haussmann and as they appeared in 1888.  We can see on this
          example that the Haussmann plan implied a large number of
          destruction and rebuilding: approximately $28,000$ houses were
          destroyed and $100,000$ were built \cite{Samuels:2012}
          (figure created from our data).}
	\label{fig:map_H}
\end{figure*}

In this article, we will consider the case of the evolution of the
street network of Paris over more than 200 years with a particular
focus on the 19th century, period when Paris experienced large
transformations under the guidance of Baron Haussmann \cite{Jordan}.
It would be difficult to describe the social, political, and
urbanistic importance and impact of Haussmann works in a few lines
here and we refer the interested reader to the existing abundant
literature on the subject (see \cite{Samuels:2012}, and \cite{Jordan}
and references therein). Essentially, until the middle of the 19th
century, central Paris has a medieval structure composed of many small
and crowded streets, creating congestion and, according to some
contemporaries, probably health problems. In 1852, Napoleon III
commissioned Haussmann to modernize Paris by building safer streets,
large avenues connected to the new train stations, central or symbolic
squares (such as the famous place de l'Etoile, place de la Nation and
place du Panth\'eon), improving the traffic flow and, last but not
least, the circulation of army troops. Haussmann also built modern
housing with uniform building heights, new water supply and sewer
systems, new bridges, etc (see Fig.~\ref{fig:map_H} where we show how
dramatic the impact of Haussmann transformations are). 

The case of Paris under Haussmann provides an interesting example
where changes due to central planning are very important and where a
naive modelling is bound to fail. We analyze here in detail the effect
of these planned transformations on the street network. By introducing
physical quantitative measures associated with this network, we are
able to compare the effect of the Hausmann transformation of the city
with its `natural' evolution over other periods.

By digitizing historical maps (for details on the sources used to
construct the maps, see the Methods section) into a Geographical
Information System (GIS) environment, we reconstruct the detailed road
system (including minor streets) at six different moments in time,
$t=1,2,\ldots,6$, respectively corresponding to years: $1789, 1826,
1836, 1888, 1999, 2010$.  For each time, we constructed the associated
primal graph $G_t$ (see the Methods section and
~\cite{Barthelemy:2011,Strano:2012}), i.e. the graph where the nodes
represent street junctions and the links correspond to road
segments. In particular, it is important to note that we have thus
snapshots of the street network before Haussmann works (1789-1836) and
after (1888-2010). This allows us to study quantitatively the effect
of such central planning.

\begin{figure*}
\centering
	\includegraphics[scale = 0.50]{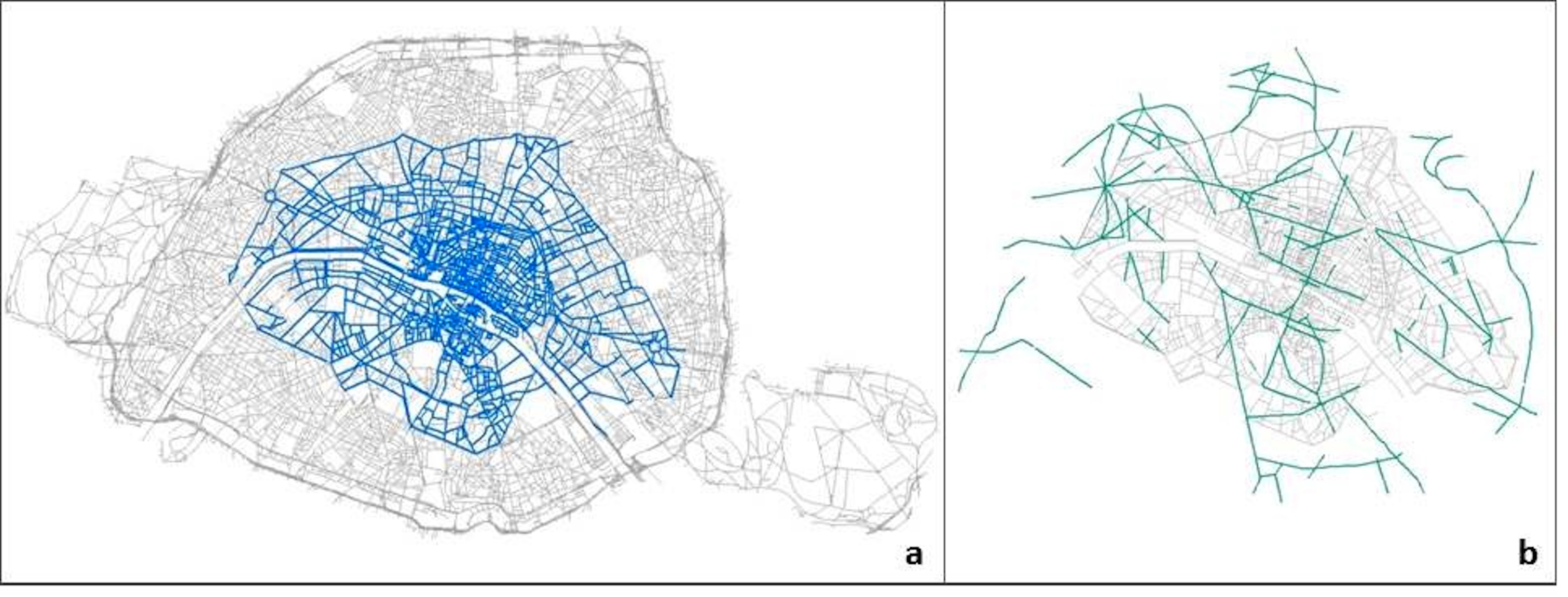}
	\caption{(a) Map of Paris in 1789 superimposed on the map of
          current 2010 Paris. In the whole study, we focus on the
          Haussmann modifications and limited ourselves
          to the 1789 portion of the street network. (b) Map of Haussmann modifications. The grey lines
          represent the road network in 1836, the green lines
          represent the Haussmann modifications which are basically
          all contained in the 1789 area (figure created from our data).}
	\label{fig:maps}
\end{figure*}

In Fig.~\ref{fig:maps}(a), we display the map of Paris as it was in
1789 on top of the current map (2010).
In order to use a single basis for comparison, we limited our study
over time to the portion corresponding to 1789. We note here that the
evolution of the outskirts and small villages in the surroundings has
certainly an impact on the evolution of Paris and even if we focus
here (mainly because of data availability reasons) on the structural
modifications of the inner structure of Paris, a study at a larger
scale will certainly be needed for capturing the whole picture of the
evolution of this city. We then have 6 maps for different times and
for the same area (of order $34km^2$). We also represent on
Fig.~\ref{fig:maps}(b), the new streets created during the Haussmann
period which covers roughly the second half of the 19th century. Even
if we observe some evolution outside of this portion, most of the
Haussmann works are comprised within this portion.

\section*{Results}

\textbf{Simple measures.} 

In the following we will study the structure of the graph $G_t$ at
different times $t$ (see the Methods section for precise definitions),
having in mind that our goal is to identify the most important
quantitative signatures of central planning during the evolution of
this road network.

\begin{figure*}
\centering
	\includegraphics[scale = 0.50]{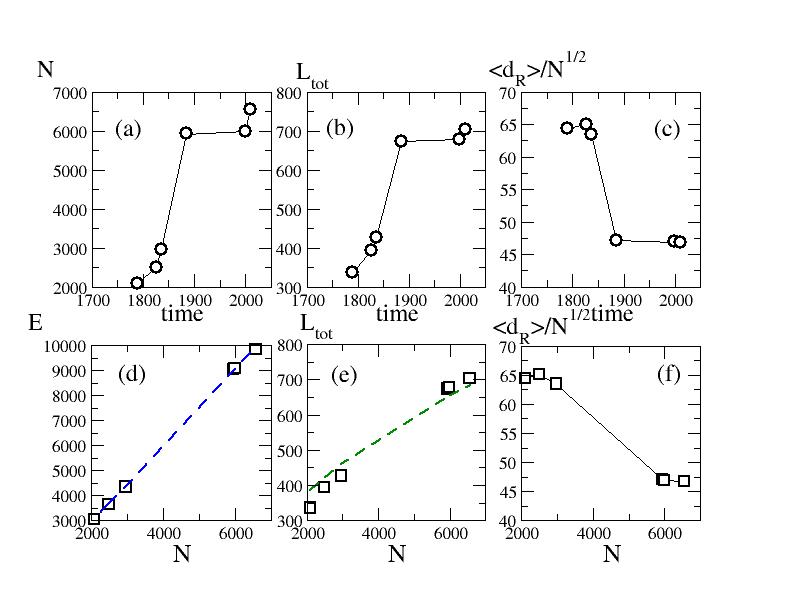}
	\caption{Top panels: Number of (a) nodes, (b) total length
          (kms), and (c) rescaled average route distance versus time.
          Bottom panels: Number of (d) edges, (e) total length (kms),
          and (f) the rescaled average route distance versus the
          number of nodes $N$. In (d) the dashed (blue) line is a
          linear fit with slope $1.55$ ($r^2=0.99$) consistent with
          constant average degree of order $\langle k\rangle\approx 3$, and
          in (e) the dashed (green) line a square root fit of the form
          $a\sqrt{N}$ with $a=8.44$kms ($r^2=0.99$). Based on a
          perturbed lattice picture this gives an area equal to
          $A\simeq 29.7$km$^2$ consistent with the actual value
          ($A=33.6$km$^2$). In (f), we show the rescaled average
          shortest route versus $N$ which decreases showing that the
          denser the network and the easier it is to navigate from one
          node to the other (if delays at junctions are neglected).}
	\label{fig:general_results}
\end{figure*}

First basic measures include the evolution of the number of nodes $N$,
edges $E$, and total length $L_{tot}$ of the networks (restricted to
the area corresponding to 1789).  In Fig.~\ref{fig:general_results} we
show the results for these indicators which display a clear
acceleration during the Haussmann period (1836-1888).
The number of nodes increased from about 3000 in 1836 to about 6000 in
1888 and the total length increase from about 400 kms to almost
700kms, all this in about 50 years. It is interesting to note that
this node increase corresponds essentially to an important increase in
the population. In particular, we note (see the Supplementary
Information for more details) that the number of nodes $N$ is
proportional to the population $P$ and that the corresponding increase
rate is of order $dN/dP\approx 0.0021$, similar to what was measured
in a previous study about a completely different area
\cite{Strano:2012}. The rapid increase of nodes during the Haussmann
period is thus largely due to demographic pressure. Now, if we want to
exclude exogeneous effects and focus on the structure of networks, we
can plot the various indicators such as the number of edges and the
total length versus the number of nodes taken as a time clock. The
results shown Fig.~\ref{fig:general_results}(d-f) display a smoother
behavior. In particular, $E$ is a linear function of $N$,
demonstrating that the average degree is essentially constant $\langle
k\rangle\approx 3.0$ since 1789. The total length versus $N$ also
displays a smooth behavior consistent with a perturbed lattice
\cite{Barthelemy:2011}. Indeed, if the segment length $\ell_1$ is
roughly constant and equal to $\ell_1=1/\sqrt{\rho}$ where $\rho=N/A$
is the density of nodes ($A$ is the area considered here), we then
obtain for the total length
\begin{equation}
L_{tot}=\frac{\langle k\rangle}{2}\sqrt{AN}
\end{equation}
A fit of the type $a\sqrt{N}$ is shown in
Fig.~\ref{fig:general_results}(d) and the value of $a$ measured gives
an estimate of the area $A\simeq 29.7km^2$, in agreement with the
actual value $A=33.6km^2$ (for the 1789 portion). This agreement
demonstrates that all the networks at different times are not far from
a perturbed lattice.

We also plot the average route distance $d_R$ defined as the average
over all pairs of nodes of the shortest route between them (see
Methods for more details). For a two dimensional spatial network, we
expect this quantity to scale as $d_R\sim \sqrt{N}$ and thus increases
with $N$. The ratio $d_R/\sqrt{N}$ is thus better suited to measure
the efficiency of the network and we observe
(Fig.~\ref{fig:general_results}(c,f)) that it decreases with time and
$N$. This result simply demonstrate that if we neglect delays at
junctions, it becomes easier to navigate in the network as it gets
denser.

\textbf{Typology of new links}


We can have three different types of new links depending on the number
of new nodes they connect. We denote by $E_i$ ($i=1,2,3$) the number
of new links appearing at time $t+1$ connecting $i$ new nodes. For
example $E_0$ counts the new links appearing at time $t+1$ connecting
two nodes existing at time $t$. In order to categorize more precisely
these new links, we use the betweenness centrality impact $\delta$
defined in \cite{Strano:2012} and which measures how a new link
(absent at time $t$ and present at time $t+1$) affects the average
betweenness centrality (see Methods section for definitions of the
betweenness centrality impact $\delta$). In \cite{Strano:2012}, the
distribution of this quantity displays two peaks which corresponds to
two types of links belonging to two distinct processes: densification
and exploration \cite{Strano:2012}. We first observe (see Figure 2 of
SI) that in the first period, the majority of new links are of the
$E_2$ type and correspond to construction of new streets with new
nodes.  We see that the Haussmann transition period (1836-1888) is not
particularly different from the other previous periods. In the modern
period (after 1999), $E_0$ becomes dominant and consistent with the
idea of a mature street network where densification dominates the
evolution of the urban tissue. Obviously, this is also an effect of
limiting ourselves to the 1789 portion: in a wider area, many new
roads were created and both densification and exploration coexist. We
note here that the structure of the street network of central Paris
remained remarkably stable from 1888 until now (and in this period
also, densification was the main process in this area).

\begin{figure*}
      \includegraphics[scale = 0.50]{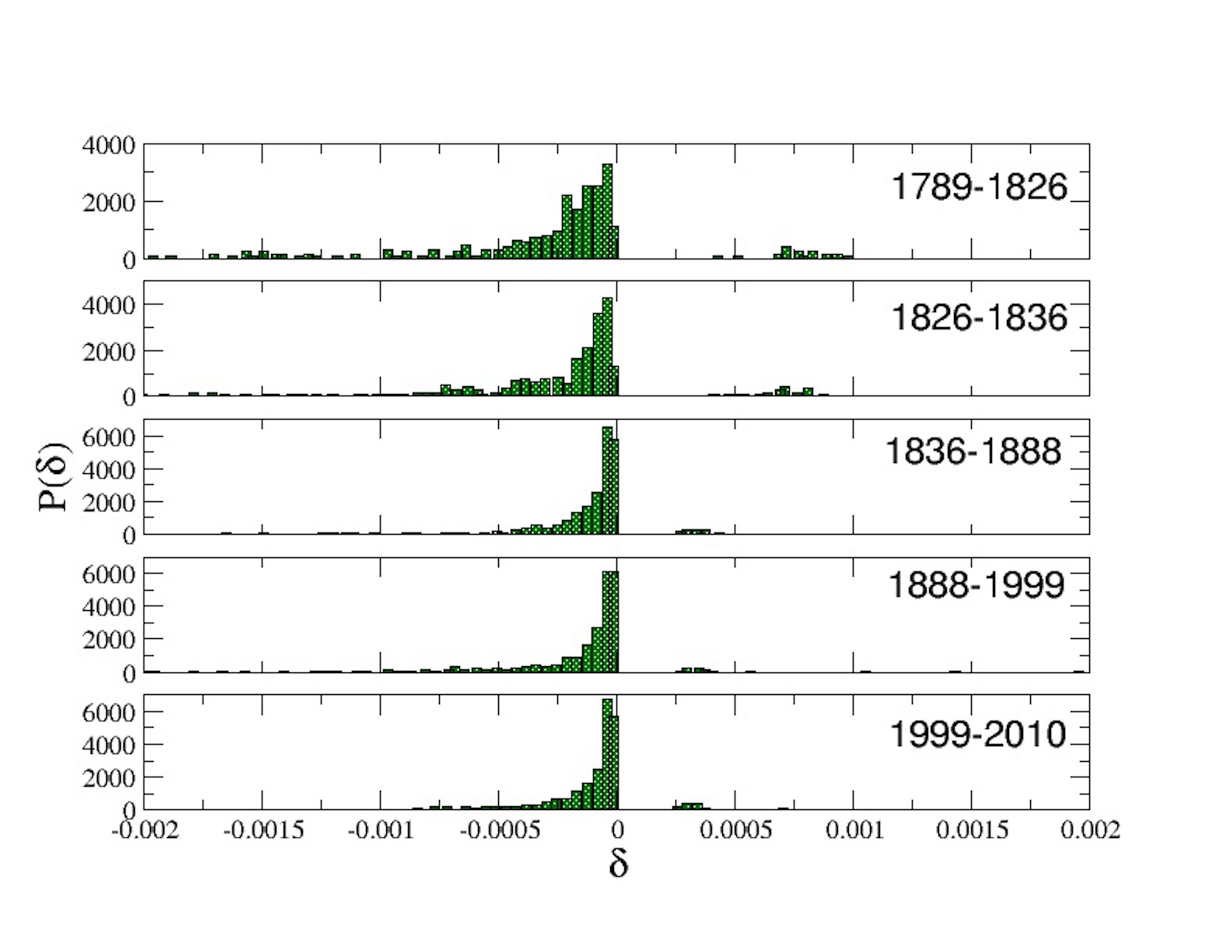}
\caption{Betweenness centrality impact distribution for the periods $1789\rightarrow
  1826$, $1826\rightarrow 1836$, $1836\rightarrow 1888$,
  $1888\rightarrow 1999$, $1999\rightarrow 2010$. This figure shows
  that densification is the main process for this portion of Paris and
  that from this point of view, the Haussmann period seems to be rather smooth and
  comparable to other periods.}
	\label{fig:bcimpact}
\end{figure*}

We then plot the distribution of this quantity $\delta$ for the
different transition periods and the result is shown in
Fig.~\ref{fig:bcimpact}. 
These figures show that for all periods most new links belong to the
densification process with a small peak of exploration in the period
1836-1888. In well-developed, mature systems, it is expected that
densification is the dominant growth mechanism. Here also, we see that
the Haussmann period is not significantly different from previous
periods.

\textbf{Evolution of the spatial distribution of centrality}

The betweenness centrality (BC) $g_v(i)$ of a node $i$ is defined in the
Methods section and essentially measures the fraction
of times a given node is used in the shortest paths connecting any pair of
nodes in the network, and is thus a measure of the contribution of a
link in the organisation of flows in the network
\cite{Freeman:1977}. In our case where we consider a limited portion of a
spatial network, two important effects need to be taken into
consideration. First, as we consider a portion, only paths within this
portion are taken into account in the calculation of the BC and this
usually does not reflect the reality of the actual origin-destination
matrix. In particular, flows with the exterior of the portion and
surrounding villages are not taken into account. As a result, the BC
will be able to detect important routes and nodes in the internal
structure of the network but will miss large-scale communication roads
such as a north-south or east-west road connecting the portion with the
surroundings of Paris. In \cite{Strano:2012}, the scale of the network
was large enough so that the BC could recover important central roads
such as Roman streets. The BC in the present case has then to be used as a
structural probe of the network, enabling us to track the important
modifications. The second point concerns the spatial
distribution of the BC which will be important in the following. For
a lattice the most central nodes (see the discussion in
\cite{Barthelemy:2011} for example) are close to the barycenter of the
nodes: spatial centrality and betweenness centrality are then usually
strongly correlated. In \cite{Lammer:2006} and \cite{Crucitti:2006} it
is shown that the most central points display interesting spatial
structures which still need to be understood, but which represent an
important signature of the networks' topology.

We first consider the time evolution of the node betweenness
centrality (with similar results for the edge BC).  In the SI (see
figure 3 of SI), we show the distribution of the node BC at different
times. Apart from the fact that the average BC varies, we see that
the tail of the distribution remains constant in time, showing that the
statistics of very central nodes is not modified. From this point of
view, the evolution of the road network follows a smooth behavior,
even in the Haussmann period.

\begin{figure*} 
	\includegraphics[scale = 0.70]{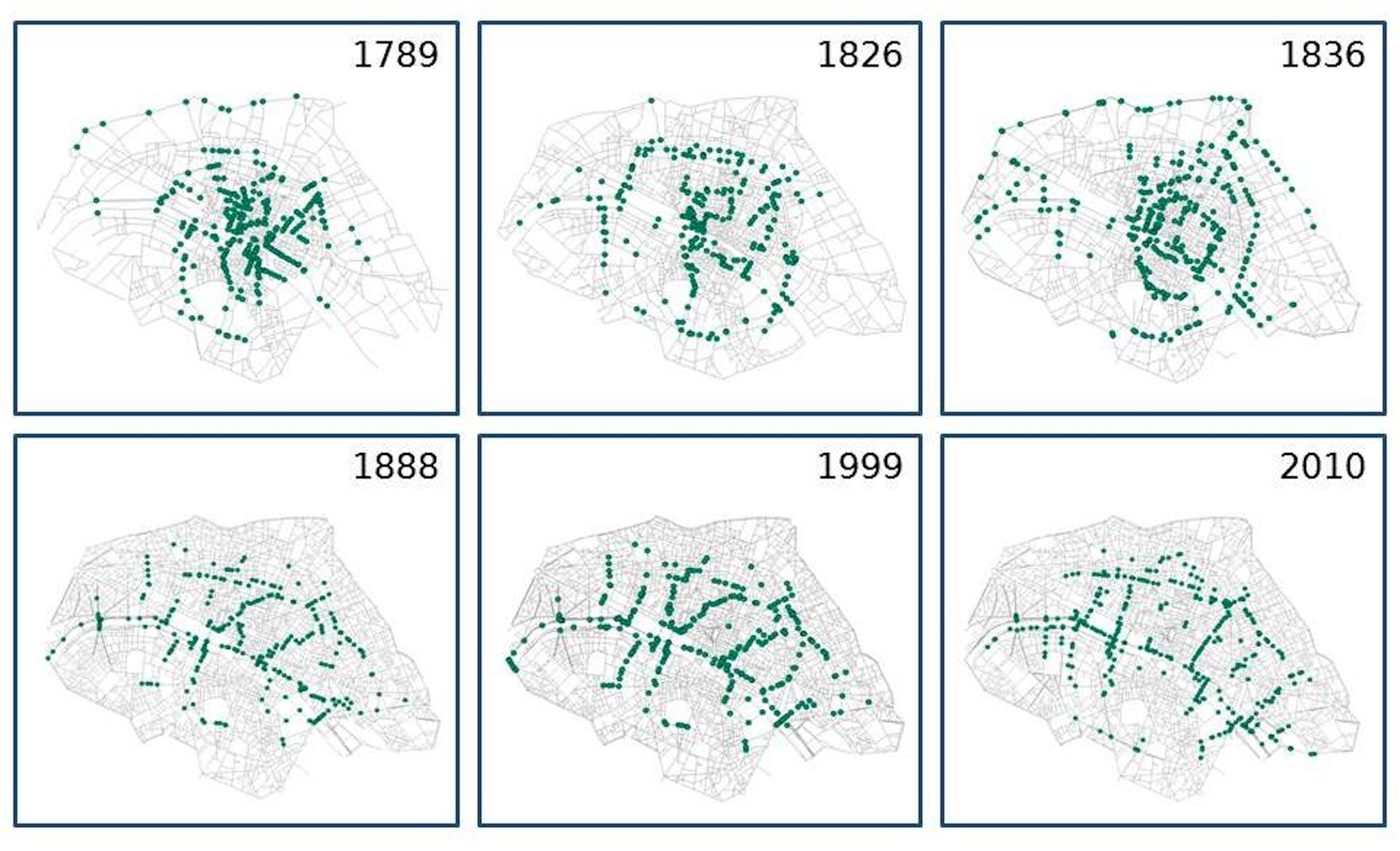}
  \caption{Spatial distribution of the most central nodes (with
    centrality $g_v$ such that $g_v>\max{g_v}/10$). We observe for the
    different periods important reorganizations of the spatial
    distribution of centrality, corresponding to different
    specific interventions. In particular, we observe a very important
    redistribution of centrality during the Haussmann period with the
    appearance of a reticulated structure on the 1888 map.}
	\label{fig:maxvbc}
\end{figure*}

So far, most of the measures indicate that the evolution of the street
network follows simple densification and exploration rules and is very
similar to other areas studied \cite{Strano:2012}. At this point, it
appears that Haussmann works didn't change radically the structure of
the city. However, we can suspect that Haussmann's impact is very
important on congestion and traffic and should therefore be seen on
the spatial distribution of centrality. In the figure
\ref{fig:maxvbc}, we show the maps of Paris at different times and we
indicate the most central nodes (such that their centrality $g_v(i)$
is larger than $\max{g_v}/\alpha$ with $\alpha=10$ \- see the SI, for
a discussion on the effect of the value of $\alpha$). 
We can clearly see here that the spatial distribution of the BC is not
stable, displays large variations, and is not uniformly distributed
over the Paris area (we represented here the node centrality, and
similar results are obtained for the edge centrality, see the SI for
plots for the edge centrality and more details). In particular, we see
that between 1836 and 1888, the Haussmann works had a dramatical
impact on the spatial structure of the centrality, especially near the
heart of Paris. Central roads usually persist in time
\cite{Mouton:1989,Strano:2012}, but in our case, the Haussmann
reorganization was acting precisely at this level by redistributing
the shortest paths which had certainly an impact on congestion inside
the city. After Haussmann we observe a large stability of the network
until nowadays.

It is interesting to note that these maps also provide details about
the evolution of the road network of Paris during other periods which
seems to reflect what happened in reality and which we can relate to
specific local interventions. For example, in the period 1789-1826
between the French Revolution and the Napoleonic empire, the maps
shown in Fig.~\ref{fig:maxvbc} display large variations with
redistribution of central nodes which probably reflects the fact that
many religious and aristocratic domains and properties were sold and
divided in order to create new houses and new roads, improving
congestion inside Paris. During the period 1826-1836 which corresponds
roughly to the beginning of the the July Monarchy, the maps in
Fig.~\ref{fig:maxvbc} suggests an important reorganization on the east
side of Paris. This seems to correspond very well to the creation
during that period of a new channel in this area (the channel `Saint
Martin') which triggered many transformations in the eastern part of
the network.

\begin{figure*} 
	\includegraphics[scale = 0.60]{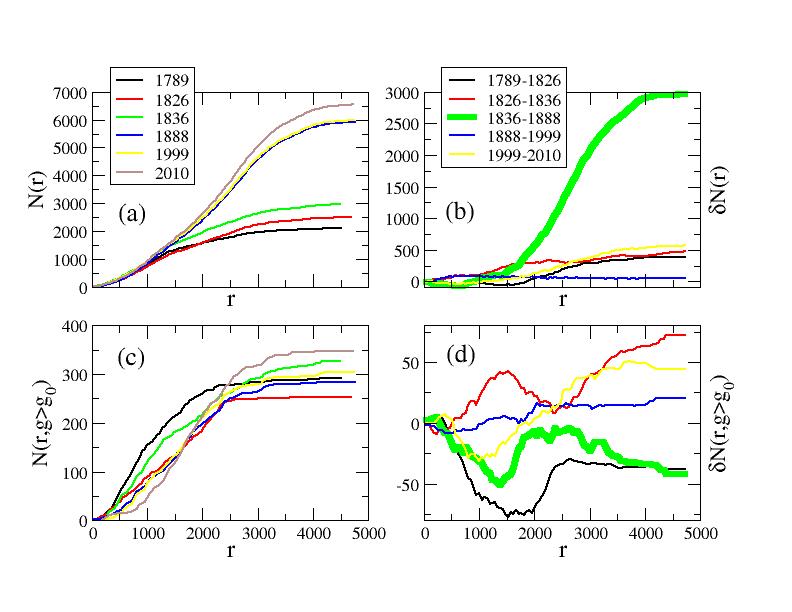}
  \caption{Top panels: (a) number of nodes in a disk of radius $r$
    from the barycenter of Paris and (b) its variation versus $r$. As
    expected the largest variation occurred during the Haussmann
    period. (c) Number of nodes at distance $r$ and with centrality larger than
    $g_0$ ($g_0=\max{g_v}/10$) and (d) its variation. The thick green line
    in the right panels indicate the Haussmann transition
    1836-1888. We see here that during the Haussmann period (and also
    in the 1789-1826 period), there is a large decrease of the number
    of central nodes in the central region of Paris ($r<2,000$ meters).}
	\label{fig:ripley}
\end{figure*}

In order to analyse the spatial redistribution effect more
quantitatively, we compute various quantities inside a disk of radius
$r$ centered on the barycenter of all nodes (which stays approximately
at the same location in time). We first study the number of nodes
$N(r)$ (Fig. \ref{fig:ripley}), its variation $\delta N(r)$ between
$t$ and $t+1$, and the number of central nodes (such that
$g_v(i)>\max{g_v}/10$). 
We see that the largest variation of the number of nodes (see
\ref{fig:ripley}(b)) is indeed in the Haussmann period 1836-1888,
especially for distance $r>1,500$ meters. More interesting, is the
variation of the most central nodes (Fig. \ref{fig:ripley}d). In particular, we observe that
during the pre-Haussmann period, even if in the period 1789-1826 there
was an improvement of centrality concentration, there is an
accumulation of central nodes both at short distances ($r<2,500$
meters) and at long distances ($r>2,500$ meters) in the following
period (1826-1836). As a result, visually clear in
Fig.~\ref{fig:maxvbc}, there is a large concentration of centrality in
the center of Paris until 1836 at least. The natural consequence of
this concentration is that the center of Paris was very probably very
congested at that time. In this respect, what happens under the
Haussmann supervision is natural as he acts on the spatial
organization of centrality. We see indeed that in 1888, the most
central nodes form a more reticulated structure excluding
concentration of centrality. A structure which remained stable until
now. Interestingly, we note that Haussmann's new roads and avenues
represent approximately $6\%$ of the total length only (compared to
nowadays network), which is a small fraction, considered that it has a
very important impact on the centrality spatial organization.

\begin{figure}
	\includegraphics[scale = 0.30]{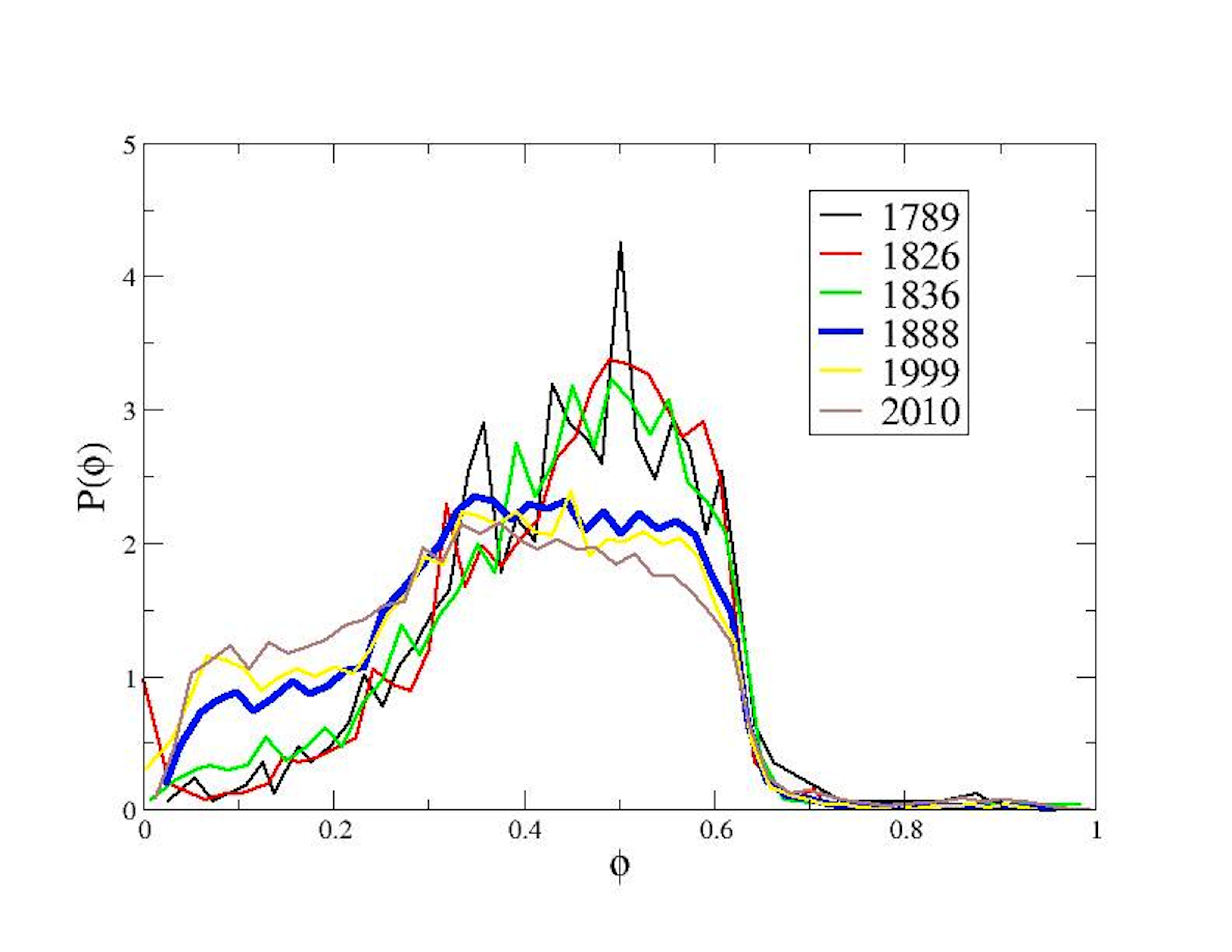}
	\caption{Probability distribution of the $\phi$ shape factor
          for the blocks at different years. Until 1836, this distribution is
          stable and we observe a dramatical change during the
          Haussmann period with a larger abundance of blocks with
          small value of $\phi$. These small values correspond to
          elongated rectangle or triangles created by streets crossing
          the existing geometry at various angles.}
	\label{fig:phi}
\end{figure}

\begin{figure}[ht!]
	\includegraphics[scale = 0.40]{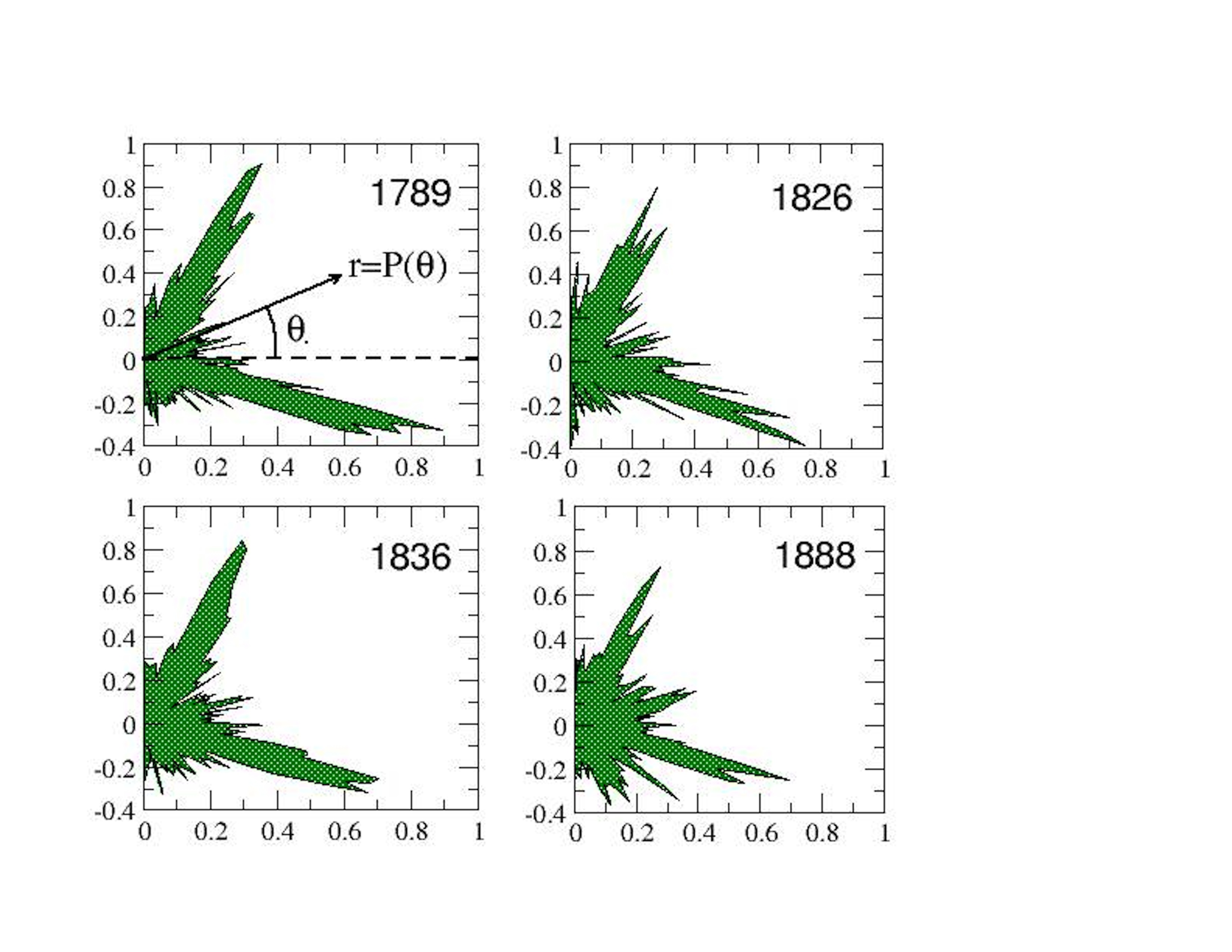}
  \caption{Radial representation of the angle distribution of road
    segments for 1789, 1826, 1836, 1888. The radial distance $r$ in
    this plot represents the probability to observe a street with
    angle $\theta$: $r=P(\theta)$ with $\theta\in [-\pi/2,\pi/2]$ and
    $P(\theta)$ is the probability to observe an oriented road with
    angle $\theta$ with the horizontal line (see first panel, top
    left). Until 1836, the distribution is peaked around two values
    separated by approximately $90$ degrees and in 1888, we observe an important
    fraction of diagonals and other lines at intermediate angles.}
	\label{fig:angles}
\end{figure}

This reorganization of centrality was undertaken with creation of new
roads and avenues destroying parts of the original pattern (see
Fig. \ref{fig:map_H} and Fig.~\ref{fig:maps}(b)) resulting in the modification of the
geometrical structure of blocks (defined here as the faces of the
planar street network). The effect of Haussmann modifications on the
geometrical structure of blocks can be quantitatively measured by the
distribution of the shape factor $\phi$ (see Methods) shown in
Fig.~\ref{fig:phi}. 
We see that before the Haussmann modifications, the distribution of
$\phi$ is stable and is essentially centered around $\phi=0.5$ which
corresponds to rectangles. From 1888, the distribution is however much
flatter showing a larger diversity of shapes. In particular, we see
that for small values of $\phi<0.25$ there is an important increase of
$P(\phi)$ demonstrating an abundance of elongated shapes (triangles
and rectangles mostly) created by Haussmann's works. These effects can
be confirmed by observing the angle distribution of roads shown on
Fig.~\ref{fig:angles} where we represent on a polar plot
$r(\theta)=P(\theta)$ with $P(\theta)$ the probability that a road
segment makes an angle $\theta$ with the horizontal line. 
Before Haussmann's modifications, the distribution has two clear peaks
corresponding to perpendicular streets and in 1888 we indeed observe a
more uniform distribution with a large proportion of various angles
such as diagonals.

\section*{Discussion}

In this paper, we have studied the evolution of the street network of
the city of Paris. This case is particularly interesting as Paris
experienced large modifications in the 19th century (the Haussmann
period) allowing us to try to quantity the effect of central
planning. Our results for central Paris reveal that most indicators
follow a smooth evolution, dominated by a densification process,
despite the important perturbation that happened during Haussmann. In
our results, the important quantitative signature of central planning
is the spatial reorganization of the most central nodes, in contrast
with other regions where self-organization dominated and which didn't
experience such a large-scale structure modification.  This structural
reorganization was obtained by the creation at a large scale of new
roads and avenues (and the destruction of older roads) which do not
follow the constraints of the existing geometry. These new roads do
not follow the densification/exploration process but appear at various
angles and intersect with many other existing roads.

While the natural, self-organized evolution of roads seems in general
to be local in space, the Haussmann modifications happen during a
relatively short time and at a large spatial scale by connecting
important nodes which are far away in the network. Following the
Haussmann interventions, the natural processes take over on the
modified substrate. It is unclear at this stage if Haussmann
modifications were optimal and more importantly, if they were at a
certain point inevitable and would have happened anyway (due to the
high level of congestion for example). More work, with more data on a
larger spatial scale are probably needed to study these important
questions.

\section*{Methods}

\textbf{Temporal Network Data} 

We denote by $G_t \equiv G(V_t, E_t)$ the obtained primal
graph at time $t$, where $V_t$ and $E_t$ are respectively the set of
nodes and links at time $t$.  The number of nodes at time $t$ is then
$N(t)= |V_t|$ and the number of links is $E(t) = |E_t|$. Using common
definitions, we thus have $V_t = V_{t-1} \cup \Delta V_t$ and $E_t =
E_{t-1} \cup \Delta E_t$, where $\Delta V_t$ and $\Delta E_t$ are
respectively the new street junctions and the new streets added in
time $]t-1,t]$ to the network existing at time $t-1$.

The networks for 1789, 1826, 1836, 1888 are extracted from the 
following maps:
\begin{itemize}
\item{} 1789: Map of the city of Paris with its new
  enclosure. Geometrically based on the `meridienne de l'Observatoire'
  and surveyed by Edm\'e Verniquet. Achieved in 1791.
\item{} 1826: Road map of Paris surveyed by Charles Picquet,
  geographer for the King and the duke of Orl\'eans.
\item{} 1836: Cadastre of Paris, Philibert Vasserot. Map constructed
  according blocks and classified according to old districts. 24 Atlas, 1810-1836.
\item{} 1888: Atlas of the 20 districts of Paris, surveyed by
  M. Alphand, and L. Fauve, under  the administration of the prefect E. Poubelle,
  Paris, 1888.
\end{itemize}
All these maps were digitized at the LaD\'eHiS under the supervision
of Maurizio Gribaudi, in the framework of a research on the social and
architectural transformations of parisian neighborhoods between the
18th and 19th centuries. The network (and the block structure of
figure 1) extracted from the Vasserot cadastre was initiated by
Anne-Laure Bethe for the program Alpage \cite{Alpage}.

The networks of 1999 and 2010 are coming from the french Geographical
National Institute (IGN) on the basis of modern surveys.

\textbf{Average route distance}

For a network, the shortest path between two nodes
is defined as the path with the minimum number of links connecting the
two nodes. For spatial networks, it makes more sense to weight the
links with their length: to each edge $e$ we thus associate a weight
given by its euclidean length $d_e$. We can then compute the length $\ell$ of a path $P$
\begin{equation}
\ell(P)=\sum_{e\in P}d_e
\end{equation}
The shortest weighted path is then the one with the minimum total length.
The average shortest weighted path is also called the average route
distance $d_R$. It indicates on average how many kilometers you have
to walk from one point to the other in this spatial network. For a two
dimensional network, it is expected \cite{Barthelemy:2011} that it
scales as
\begin{equation}
d_R\sim N^{1/2}
\end{equation}
for a network of size $N$. In order to compare networks with different
numbers of nodes $N$, it is then natural to compare the rescaled
average route distance $d_R/\sqrt{N}$.

\textbf{Betweenness centrality, Impact}
 
The nature of the growth process can be quantitatively characterised
by looking at the centrality of streets. Among the various centrality
indices available for spatial networks we use here the betweenness
centrality (BC) \cite{Freeman:1977,Crucitti:2006,Porta:2006}, which is
one of the measures of centrality commonly adopted to quantify the
importance of a node or a link in a graph. Given the graph $G_t\equiv
G(V_t,E_t)$ at time $t$, the BC of a link $e$ is defined as: 
\begin{equation}
  g(e) = \sum_{i\in V} \sum_{\stackrel{j \in V} {j \neq i}} 
\frac{\sigma_{ij}(e)}{\sigma_{ij}}
\end{equation}
where $\sigma_{ij}$ is the number of shortest paths from node $i$ to
node $j$, while $\sigma_{ij}(e)$ is the number of such shortest paths
which contain the link $e$. The quantity $g(e)$ essentially measures
the number of times a link is used in the shortest paths connecting
any pair of nodes in the network, and is thus a measure of the
contribution of a link in the organisation of flows in the network. 
The BC of a node is defined in a similar way
\begin{equation}
g_v(i)=\sum_{s,t\in V}\frac{\sigma_{st}(i)}{\sigma_{st}}
\end{equation}
where $\sigma_{st}(i)$ denotes here the number of shortest path from
node $s$ to $t$ going through the node $i$.

In order to evaluate the impact of a new link on the overall
distribution of the betweenness centrality we use the betweenness
centrality impact defined in \cite{Strano:2012}. In the graph at time $t$,
we first compute the average betweenness centrality of all the links
of $G_t$ as:
\begin{equation}
\overline{g}(G_t) = \frac{1}{(N(t) - 1) (N(t) - 2)} \sum_{e\in E_t} g(e)
\end{equation}
where $g(e)$ is the betweenness centrality of the edge $e$ in the
graph $G_t$. Then, for each link $e^* \in \Delta E_t$, i.e. for each
newly added link in the time window $]t-1,t]$ we consider the new
graph obtained by removing the link $e^*$ from $G_t$ and we denote
this graph as $G_t \setminus \{ e^* \}$.  We compute again the average
edge betweenness centrality, this time for the graph $G_t \setminus \{
e^* \}$. Finally, the impact $\delta(e^*)$ of edge $e^*$ on the
betweenness centrality of the network at time $t$ is defined as
\begin{equation}
\delta(e^*) = \frac { \left[\overline{g}(G_t) -  \overline{g}( G_t \setminus \{ e^* \} ) \right] }{
 \overline{g}(G_t)}
\end{equation}
The BC impact is thus the relative variation of the graph average
betweenness due to the removal of the link $e^*$.

\textbf{Form factor}

The shape or form factor $\phi$ of blocks is defined as the ratio of
the area of the block and the area of the circumscribed circle of
diameter $D$ (see \cite{Lammer:2006,Barthelemy:2011})
\begin{equation}
\phi=\frac{A}{\pi D^2/4}
\end{equation}
The more anisotropic the block and the smaller the factor $\phi$.

\section*{Supplementary Information}

\subsection{Population and nodes}

\begin{figure}
	\includegraphics[scale = 0.30]{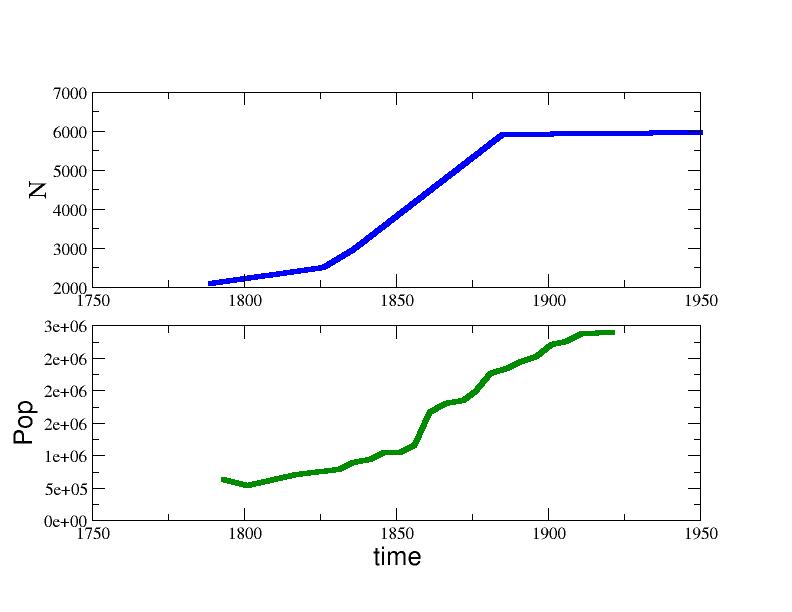}
	\includegraphics[scale = 0.30]{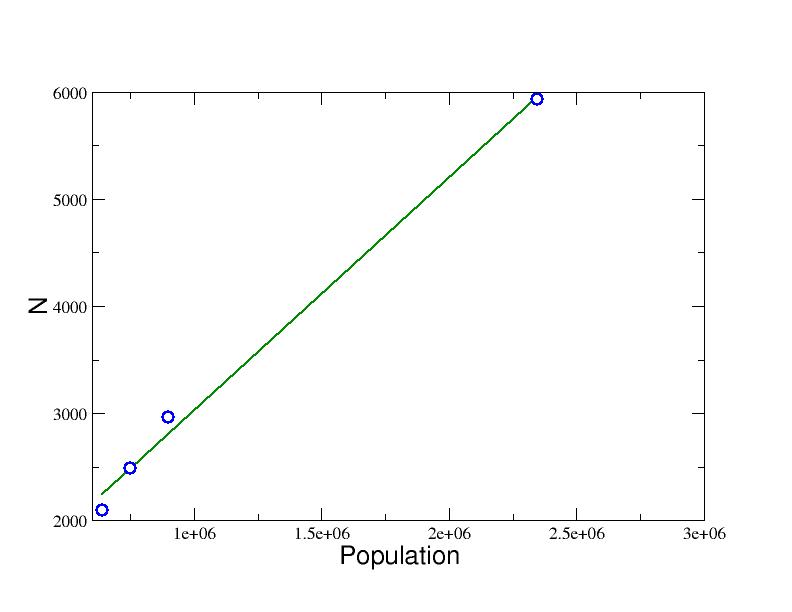}
\caption{Top panel: Evolution of the number of nodes versus time for
  Paris. Middle panel: evolution of the Paris population. Bottom
  panel: Number of nodes versus population. The line is a linear fit ($r^2>0.99$).}
	\label{fig:popu}
\end{figure}

In figure \ref{fig:popu}, we show the evolution of the number of nodes and of the
population of Paris (for the 12 districts delimited by the `fermiers generaux'
for the period 1789-1851 and after for the 20th districts of Paris).
The area under consideration for the calculation of the population is
not exactly the same, and only the order of magnitude can be trusted
here. We can compute the number of nodes $N$ versus the population $P$
and we observe a linear dependence with coefficient $dN/dP=0.0021$ (in
previous studies, we also found a linear dependence [24], but with a
linear coefficient equal to $dN/dP=0.019$). It is thus clear that the
number of nodes follows the demographic population and that the large
increase observed during the Haussmann period is largely due to the
demographic pressure.

\subsection{Type of new links}

\begin{figure}
	\includegraphics[scale = 0.30]{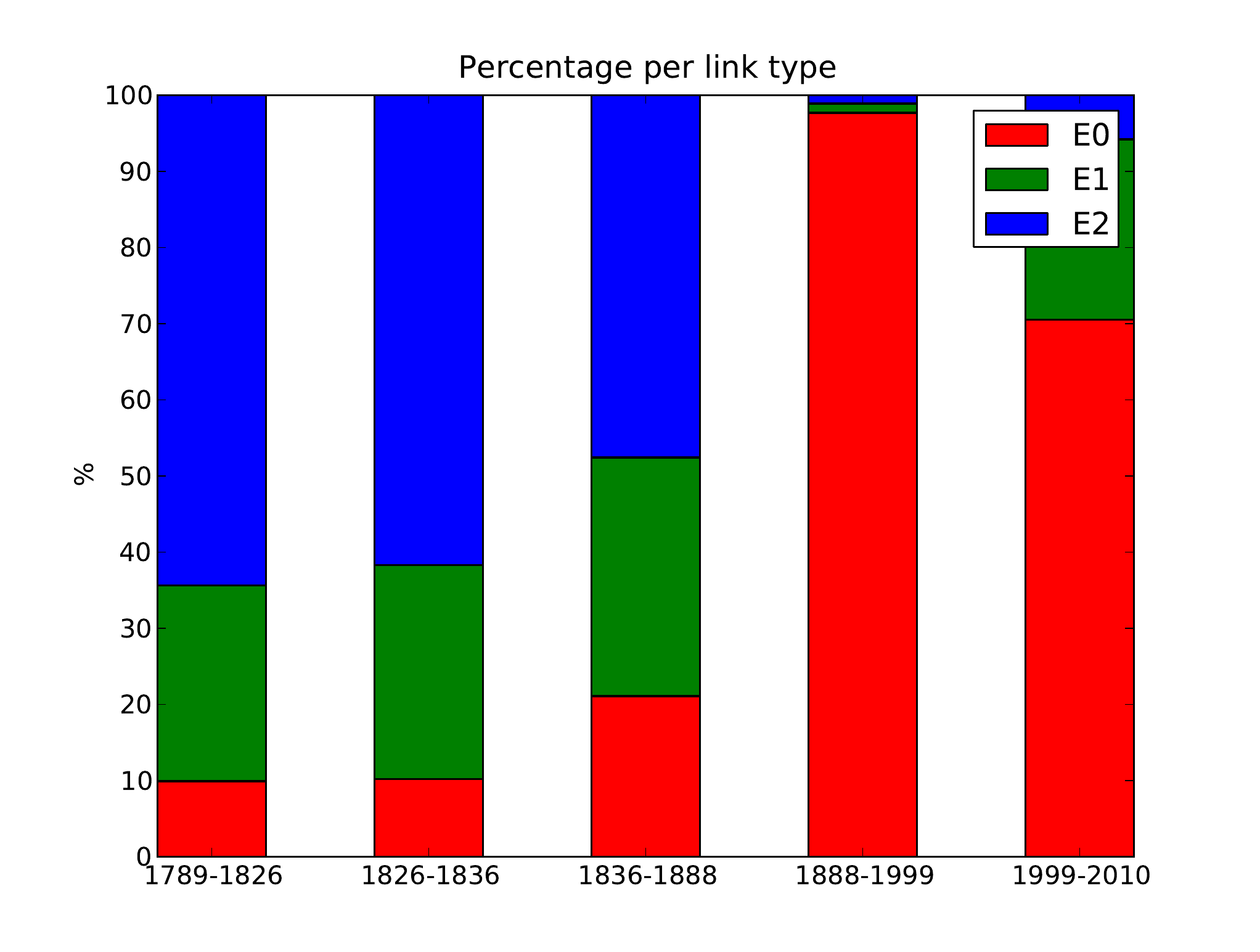}
\caption{Evolution of the percentage of different types of
  links. $E_i$ corresponds to new links with creation of $i$ new
  nodes. The denser the network, the smaller $E_2$ and in particular,
  we observe that the Haussmann period is not radically different in
  this respect from other periods.}
	\label{fig:newlinks1}
\end{figure}

In figure \ref{fig:newlinks1}, we show the evolution of the proportion of the different
types of new links. We see in this figure that the evolution is rather
smooth and that from this point of view, the Haussmann period is not
radically different from previous ones.

\subsection{Stability of the BC distribution}

\begin{figure}
	\includegraphics[scale = 0.30]{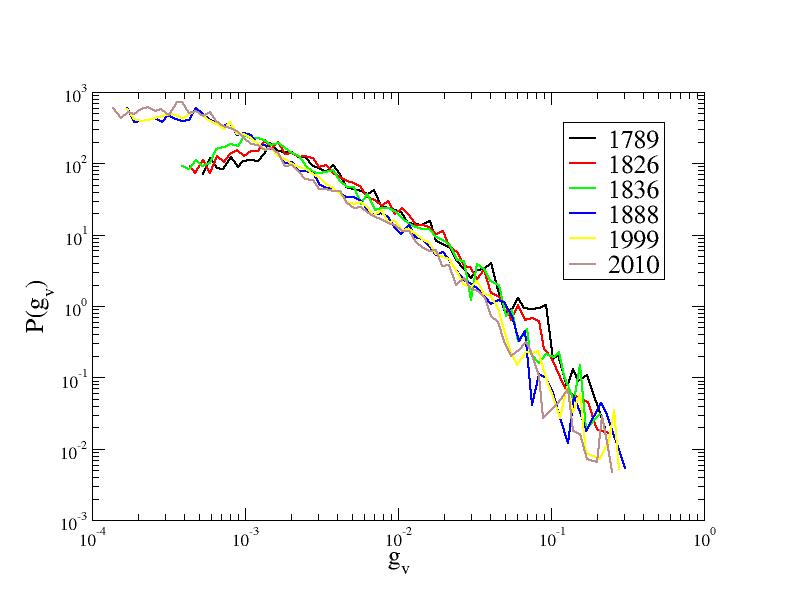}
 \caption{\textbf{Vertex BC distribution for all links and all
     periods.}  Vertex BC probability distribution for the different
          time snapshots considered in this study. We note that the
          average BC decreases (indicating a larger navigability in
          the system) and that the overall shape and tail remain the
          same across all times.}
\label{fig:SI_fig2}
\end{figure}

We consider here the evolution of the vertex BC with time. In figure
\ref{fig:SI_fig2}, we see that the average BC decreases slightly and that
the overall probability distribution remains constant in time.

\subsection{Most central nodes: stability of spatial patterns}

\begin{figure*}
	\includegraphics[scale = 0.60]{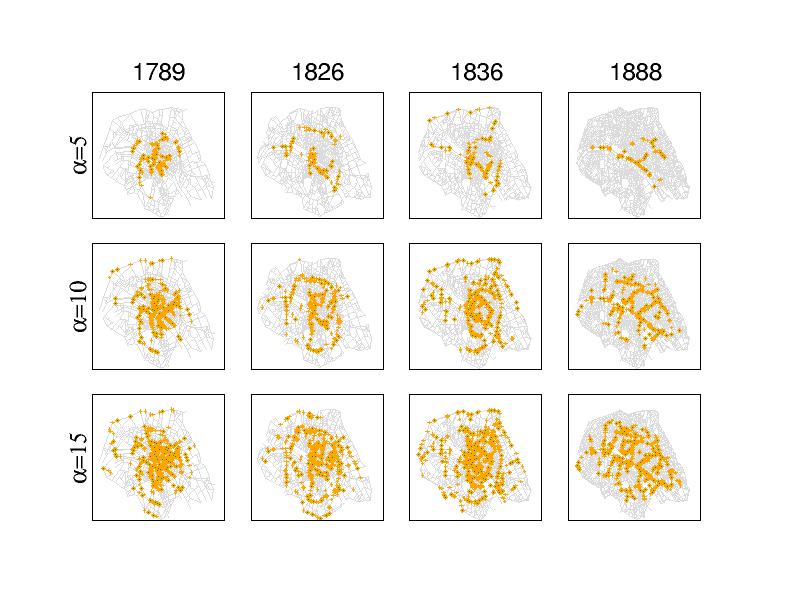}
\caption{Spatial patterns of most central points defined by
  $g_v>\max{g_v}/\alpha$ for different values of $\alpha$. We see that
  for the 4 time points considered here that the pattern is robust with
  respect to the value of $\alpha$. }
	\label{fig:alpha}
\end{figure*}

The most central nodes are such as their centrality is
$g_v>\max{g_v}/\alpha$. In the letter we consider $\alpha=10$ and we
show in figure \ref{fig:alpha} the results for $\alpha=5$ and $\alpha=15$. A visual inspection
shows that the patterns are rather robust versus $\alpha$ and that
$\alpha=10$ corresponds to an intermediate situation displaying
interesting patterns.

\subsection{Spatial pattern of the most central edges}

\begin{figure*}
	\includegraphics[scale = 0.60]{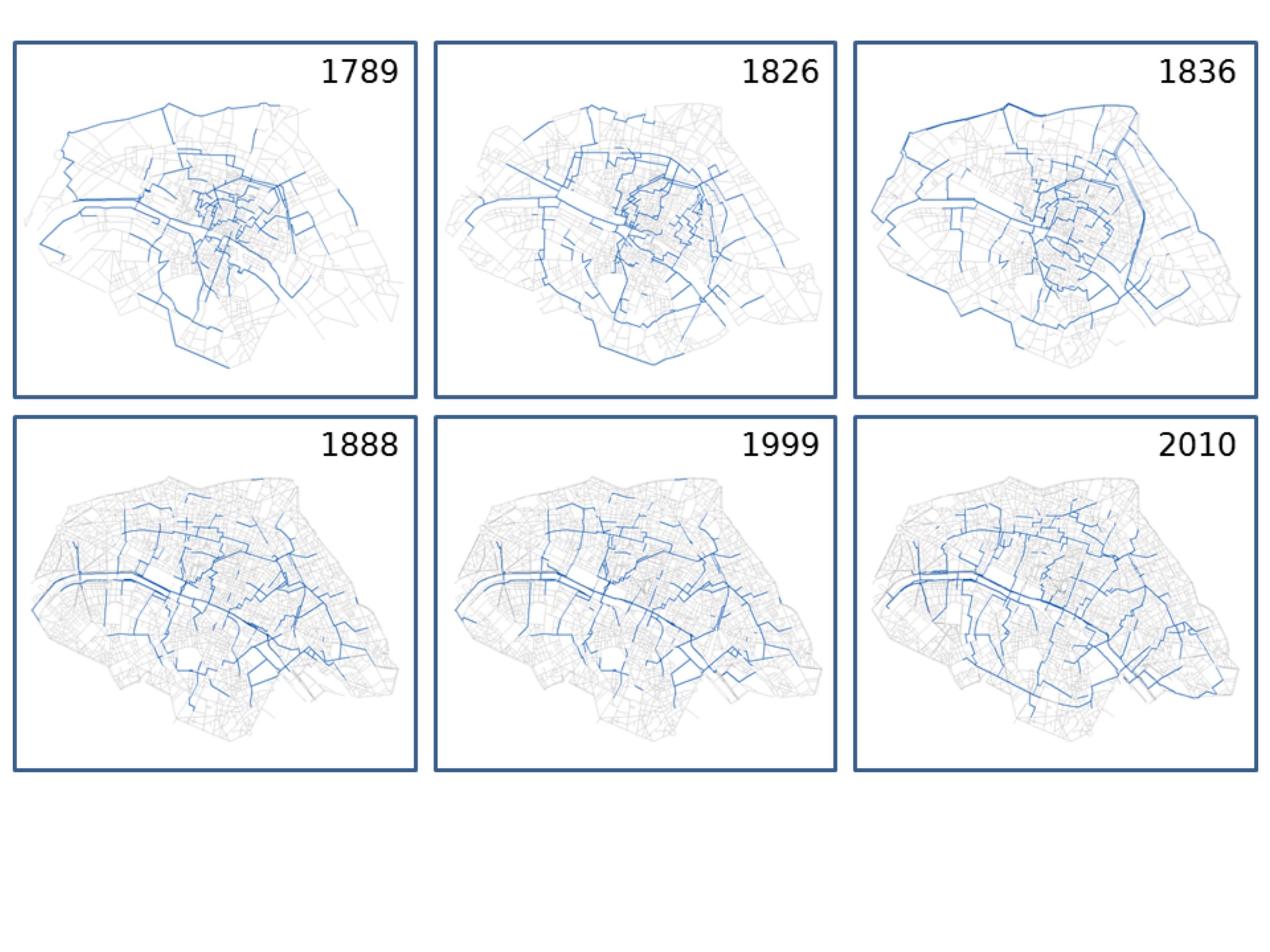}
\caption{Spatial patterns of most central edges defined by
  $g_e>\max{g_e}/\alpha$ for $\alpha=20$.}
	\label{fig:ebc}
\end{figure*}

Instead of the most central nodes, we can also represent the most
central edges such that their centrality is $g_e>\max{g_e}/\alpha$. If
we consider here $\alpha=20$ we obtain for the different dates
the results presented in Fig.~\ref{fig:ebc}.
We can see that the pattern for the edges is naturally consistent with
the one obtained with the node centrality.


{\bf Acknowledgements.} MB acknowledges funding from the EU Commission
through project EUNOIA (FP7-DG.Connect-318367). HB acknowledges
funding from the European Research Council under the European Union's
Seventh Framework Programme (FP/2007-2013) / ERC Grant Agreement
n.321186 - ReaDi -Reaction-Diffusion Equations, Propagation and
Modelling.


\end{document}